\documentstyle[preprint,aps,floats]{revtex}
\tightenlines
\input psfig.tex 
\begin{document}
\def\ba{\begin{eqnarray}}
\def\ea{\end{eqnarray}}
\def\be{\begin{equation}}
\def\ee{\end{equation}}
\def\({\left(}
\def\){\right)}
\def\[{\left[}
\def\]{\right]}
\def\lagrange {{\cal L}}
\def\del {\nabla}
\def\d {\partial}
\def\Tr{{\rm Tr}}
\def\half{{1\over 2}}
\def\fourth{{1\over 8}}
\def\bibi{\bibitem}
\def\S{{\cal S}}
\def\H{{\cal H}}
\def\K{{\cal K}}
\def\xx{\mbox{\boldmath $x$}}
\newcommand{\phpr} {\phi'}
\newcommand{\gam}{\gamma_{ij}}
\newcommand{\sqgam}{\sqrt{\gamma}}
\newcommand{\delk}{\Delta+3{\cal K}}
\newcommand{\dph}{\delta\phi}
\newcommand{\om} {\Omega}
\newcommand{\dom}{\delta^{(3)}\left(\Omega\right)}
\newcommand{\rar}{\rightarrow}
\newcommand{\Rar}{\Rightarrow}
\newcommand{\labeq}[1] {\label{eq:#1}}
\newcommand{\eqn}[1] {(\ref{eq:#1})}
\newcommand{\labfig}[1] {\label{fig:#1}}
\newcommand{\fig}[1] {\ref{fig:#1}}
\def\gsim{ \lower .75ex \hbox{$\sim$} \llap{\raise .27ex \hbox{$>$}} }
\def\lsim{ \lower .75ex \hbox{$\sim$} \llap{\raise .27ex \hbox{$<$}} }
\newcommand\bigdot[1] {\stackrel{\mbox{{\huge .}}}{#1}}
\newcommand\bigddot[1] {\stackrel{\mbox{{\huge ..}}}{#1}}

\title{Homogeneous Modes of Cosmological Instantons}
\author{
Steven Gratton\thanks{email: S.T.Gratton@damtp.cam.ac.uk}
 and Neil
Turok\thanks{email: N.G.Turok@damtp.cam.ac.uk}}
\address{
DAMTP, CMS, Wilberforce Road, Cambridge, CB3 0WA, U.K.}
\date{\today}
\maketitle

\begin{abstract}

We discuss the $O(4)$ invariant perturbation modes of
cosmological instantons.
These modes are 
spatially homogeneous in Lorentzian spacetime
and thus not relevant to 
density perturbations. But
their properties are important in establishing the
meaning of the Euclidean path integral. If negative modes
are present, 
the Euclidean path integral
is not well defined, 
but may nevertheless be useful in an approximate 
description 
of the decay of an unstable state. When gravitational dynamics
is included, counting negative modes
requires a careful treatment of the 
conformal factor problem.
We demonstrate that for an appropriate choice of 
coordinate on phase space,  
the second order Euclidean action is bounded below
for normalized perturbations and has 
a finite number of negative modes. 
We 
prove that there is a 
negative mode for many gravitational instantons
of the Hawking-Moss or Coleman-De Luccia type, and
discuss the associated spectral flow.
We also investigate Hawking-Turok constrained instantons,
which occur in a generic
inflationary model. Implementing the 
regularization and constraint 
proposed by Kirklin, Turok and Wiseman, we find that
those instantons
leading to substantial inflation do not possess 
negative modes. Using an alternate
regularization and 
constraint motivated by reduction from
five dimensions,  we find a negative mode is present.
These investigations shed new light on the suitability of
Euclidean quantum gravity as a potential description of our
universe.

\end{abstract}

\vskip .2in

\section{Introduction}

An important issue in the study of quantum
gravity is the question of 
whether a consistent Euclidean formulation exists at all.  There is of
course the problem of renormalizability, but this may be considered a
``technical'' difficulty perhaps to be resolved by the
inclusion of more degrees of freedom at high energies.
Somewhat more fundamental is the apparent 
unboundedness of the Euclidean action itself, 
known as ``the conformal factor problem''~\cite{ghp}. This problem
has a deep physical origin in the fact that
there is no canonical
ensemble for gravitating systems. This is perhaps the 
major hazard to be faced 
the use of non-perturbative Euclidean techniques.

In this paper
we study the behaviour of the action around a class of
non-perturbative $O(4)$ invariant classical 
solutions of Euclideanised Einstein-scalar field
theory called cosmological
instantons.  These instantons have been used for some
time in inflationary theory to 
describe the decay of an inflating
false vacuum state~\cite{coldel}.
In analogy with the instanton
description of quantum tunneling~\cite{col,calcol}, one expects such
instantons 
to possess 
a single negative mode~\cite{col2}, leading to an imaginary contribution 
to the energy of the unstable state.
For the tunneling interpretation to be valid, it is important 
to establish the presence of the negative mode. However, 
if such a mode exists this equally establishes that the 
Euclidean path integral
can only be regarded as an approximation
since it is ill-defined at a fundamental level.

Cosmological instantons are also used in another,
more ambitious context. They provide a first approximation
to the  Euclidean no boundary path integral~\cite{HH},
and therefore are the possible foundation 
for a description of the initial conditions of the
universe itself. One seeks solutions of the
no boundary form, in which Lorentzian spacetime
is analytically rounded off on a Euclidean region.
Correlators of observables are then to be computed 
via a perturbation expansion in the Euclidean region.
Since the Euclidean propagator is 
unique, in principle one should obtain 
unique Euclidean correlators which then,
after analytic continuation to the Lorentzian region,
fully define the theory~\cite{gt}.

The existence of negative modes should we believe be 
a major consideration in deciding whether or not
instantons should be regarded as describing tunneling 
or whether they provide a fundamental description of
the initial state for the universe. The complete set
of fluctuation modes divides into those which are
$O(4)$ invariant in the Euclidean region and those
which are not. The latter describe 
inhomogeneous cosmological perturbations, and 
it is well known that they possess positive 
Euclidean action.
Negative modes can however 
arise in the $O(4)$ invariant sector and in
this paper we shall develop the technology 
necessary to describe them.

At first sight the conformal factor problem makes the
problem of defining the number of negative modes 
intractable.
If an inappropriate choice of variables is made, as 
in~\cite{rub} for example, the Euclidean action is
unbounded below with an infinity of negative modes appearing.
Another approach has been proposed which involves
Wick rotating infinite
sub-classes of modes in the Euclidean region and arguing about the
transformation properties of the measure~\cite{sastan,tan}.  This is
related to the proposal of Gibbons, Hawking and Perry~\cite{ghp} of
Wick rotating the conformal factor fluctuations to make
the Euclidean path integral bounded.  These methods seem 
rather arbitrary and contrived, and do not seem to yield
sensible results when applied to both
Hawking-Moss as well as Coleman-De Luccia instantons.

Instead we shall attempt to continue to the
Euclidean region in a well-defined manner following from 
a Hamiltonian formulation in the Lorentzian region. 
We integrate out gauge degrees of freedom
in the Lorentzian region and 
analytically continue only physical degrees of freedom.
It is important to note that the 
analytic continuation which generalizes the choice $t \rightarrow -i \tau$ 
in Minkowski spacetime is completely
fixed by considering a hypothetical field only weakly coupled to
gravity, and demanding that its action be positive definite in
the Euclidean region. 

To quadratic order in the fluctuations, 
``three--fourths'' of the 
conformal factor problem is solved by removing gauge degrees
of freedom and
taking the Einstein $G_{0\mu}$ 
constraint equations properly into account.  
The latter
link the 
variation of the metric with the amplitude of the scalar field, so a
quickly oscillating metric leads to a large
scalar field fluctuation and large Euclidean action. 
This 
eliminates 
negative kinetic terms for $O(4)$ non-invariant 
modes (i.e. the spatially inhomogeneous modes in the Lorentzian
universe), as discussed for example in Ref.~\cite{gt}.

The remaining negative kinetic terms are associated with
the $O(4)$ invariant modes. As discussed above, the negativity
is meaningful and we should not attempt to artificially remove it.
Rather we seek to isolate it in a discrete number of
clearly identified fluctuation modes. Changes of variable
in the path integral can be very helpful for this purpose.
Consider for example quantum mechanics in 
real time for a particle with a positive harmonic potential
but a negative kinetic term. The real time path integral may be written
\be
\int \left[ dq\right] e^{i \int dt (-{A \over 2} \dot{q}^2 -{B\over 2}
q^2)}.
\labeq{f1}
\ee
where both $A$ and $B$ are positive. 
If we perform the usual analytic continuation $t=-i\tau$, we obtain 
a Euclidean action with a negative kinetic term, analogous to
the case of gravity. For normalized fluctuations
in $q$ the Euclidean action possesses an infinite number of
negative modes. However a functional Fourier transform 
sheds new light on the problem. We can reproduce the 
first term in the action is with
a functional integral over $p$, with action $\int p \dot{q} +{1 \over
  2A} p^2$.
We then integrate by parts in $t$, and functionally integrate over
$q$, obtaining in place of~\eqn{f1}
\be
\int \left[ dp\right] e^{i \int dt ({1\over 2B} \dot{p}^2 +{1\over 2A } 
p^2)}.
\ee
We notice that the coefficients of the kinetic and potential terms
have been interchanged, so that we now have a positive 
kinetic term and a negative potential term. 
Continuing as before via $t=-i\tau$, the Euclidean
action now has positive kinetic term. Of course the 
potential term
is now negative, so 
we have merely replaced one ill-defined Euclidean
path integral with another and one might think we had not
gained much. But for the context below,
where the Euclidean region is compact, the positivity
of the kinetic term means that even if the potential term
is negative, the action is bounded below for
normalized perturbations and the 
number of negative modes is then finite. 

The functional Fourier transform used above is of
course just another way of introducing the
full first order on phase space $(p,q)$ 
appropriate to a Hamiltonian treatment. 
The freedom we then exploit is the choice of the linear combination
of $p$ and $q$ for the retained variable to be
continued to the Euclidean region.
If some particular linear combination
yields a positive kinetic term throughout the Euclidean 
region, this is to be preferred since the number of
negative modes is then countable.
Note that just one 
linear combination of $p$ and $q$ is sufficient 
to completely define the theory in the Euclidean region,
since correlators of the independent linear combination
may be derived by differentiating with respect to time
and using 
the Heisenberg equations of motion
($q= -\dot{p}/B$ in the example above).

Using the freedom to define the retained coordinate,
and exploiting the fact
that the number of negative modes of a Sturm-Liouville operator is
independent of the measure chosen, 
we prove that large
classes of regular gravitational instantons
have negative modes. 
This puts tunneling interpretations of
these instantons on a firmer footing.  But
as discussed above it raises doubts about using them
to describe the beginning of the universe.

Indeed in theories with Hawking-Moss and Coleman-De Luccia
instantons, there is usually a lower action instanton 
which does not possess negative modes. Consider
theories where there is a global potential minimum, 
and it is positive. Then there is an instanton which is
a round four sphere. The radius of the sphere 
tends to infinity as the potential 
minimum decreases to zero. This instanton solution
has no negative modes. Its analytic continuation
is just empty de Sitter spacetime, or in the limit of
zero potential minimum, Minkowski spacetime.
It seems to us that this trivial vacuum state, defined by 
the lowest action instanton is
in fact the natural one implied by the Euclidean no boundary
proposal for the sector of the theory with 
the simplest, $S^4$ topology. Selecting 
another instanton with this topology
(Hawking-Moss, or Coleman-De Luccia) to describe 
the beginning of the universe seems 
unacceptable. Since those 
instantons possess negative modes, they may
describe tunneling from one approximate, unstable state 
but to use them 
as the basis for a fundamental description is surely questionable.

The existence of singular, but finite action, constrained
instanton solutions~\cite{HT} in a generic inflationary model
opens new possibilities in this regard. Such 
instantons
may be made regular by a change of variables on superspace
plus an appropriate regularization of the potential 
$V(\phi)$ at large values of the inflaton field 
$\phi$~\cite{ktw}. In the regular description, the topology of the
solutions is not $S^4$ but  $RP^4$, and the scalar
field is actually a twisted field living on that manifold.
These instantons are classical solutions in a sector of the theory
which is topologically distinct from the naive $S^4$
Euclidean vacuum discussed above.
Implementing the 
regularization scheme of 
Kirklin, Turok and Wiseman~\cite{ktw}, we show 
the constraint removes  
negative modes for those instantons giving substantial
inflation. There is therefore a stable valley in
the configuration space of the Euclidean theory,
and the 
Euclidean path integral for
fluctuations 
about such solutions is well defined to quadratic order.
It 
may therefore possess a well defined 
perturbative
expansion to higher orders.

The outline of the paper is as follows. 
We review negative modes of Hawking-Moss
instantons, before discussing regular Coleman-De Luccia
instantons and then generic singular Hawking-Turok instantons.  
Motivated by the construction of Ref.~\cite{ktw} we regularize 
the latter by matching the scalar potential $V(\phi)$ at large 
$\phi$ to a
certain class of exponentially decaying
potentials.
We discuss an alternate regularization and constraint 
motivated by Garriga'a construction of singular instantons 
as dimensionally reduced five dimensional regular solutions~\cite{garriga}. 
In the latter construction, a negative mode is always 
present for a generic slow-roll potential. 
Finally 
implications of this work for Euclidean quantum gravity are commented
upon.  

Our study yields a simple picture of
Euclidean configuration space for a generic inflationary theory,
into which the known classical solutions fit.
The valley we have identified for
Hawking-Turok instantons is potentially of much
interest since it may provide a well defined 
perturbative basis for Euclidean approaches to inflationary cosmology. 

We would like to draw to the attention of the reader the
recent work of A. Khvedelidze, G. Lavrelashvili and 
T. Tanaka~\cite{lav2,lav1}, which also addresses the issue of negative
modes about Coleman-De Luccia instantons.

\section{The Second Order Action}

Our starting point is the second order
action for scalar perturbations in the Lorentzian universe, as
discussed in Sec. 4 of~\cite{gt}. We consider a scalar
field $\phi$ with potential $V(\phi)$ minimally coupled to gravity. 
The background field equations are 

\be
\phi''+2{\cal H} \phi' =-a^2 V_{,\phi}(\phi)
\qquad {\cal H}^2 = {\kappa\over 3} ({1\over 2}
\phi'^2 +V(\phi) a^2) -{\cal K}, 
\ee

where $\kappa=8\pi G$, ${\cal H}= a'/a$, primes 
denote derivatives with respect to conformal time  and ${\cal K}=0,\pm1$
for flat, closed and open FRW universes.
With the perturbed line element

\ba
ds^2=a^2\(-\(1+2A\) d\tau^2 + B_{|i} dx^i d\tau + \( \(1-2\psi\)
\gamma_{ij}+ 2 E_{|ij}\) dx^i dx^j \),
\ea

and the scalar field represented as $\phi+\delta \phi$, with 
$\phi$ the background solution,
the second order action for fluctuations is given by Eq.~(18)
of~\cite{gt}, reproduced here: 

\ba S_2 & = & \frac{1}{2\kappa} \int d\tau d^3 x a^2 \sqrt{\gamma} \bigg\{
  -6\psi'^2-12{\cal H} A\psi'+2\Delta\psi\(2A-\psi\)-2\({\cal
    H}'+2{\cal H}^2\)A^2
 \nonumber \\
& & +\kappa\(\dph'^2+\dph\Delta\dph-a^2 V_{,\phi\phi}\dph^2\)
+2\kappa\(3\phpr\psi'\dph-\phpr\dph'A-a^2V_{,\phi}A\dph\) \nonumber \\
& & +{\cal K}\(-6\psi^2+2A^2+12\psi A + 2\(B-E'\)\Delta\(B-E'\)\) \nonumber \\
& &  +4\Delta\(B-E'\) (\frac{\kappa}{2} \phpr\dph-\psi'-{\cal
    H}A) \bigg\}.  \labeq{2acnew} \ea
  This is well-defined for all values of $\phpr$ and the three-space
  Laplacian $\Delta$.  In an open universe $\Delta$ takes the value
  zero for the spatially homogenous mode, and $-p^2-1$ with $p^2>0$ for
  the continuum of square integrable modes.  In a closed universe
  $\Delta$ is given by $-n^2+1$ with $n \in {\cal N}$.

The momenta canonically conjugate to $\psi$, $E$, and $\dph$ are
\ba
\Pi_\psi &=&
\frac{2a^2\sqrt{\gamma}}{\kappa}\(-3\psi'+3\frac{\kappa}{2}\phpr\dph-3\H
  A-\Delta\(B-E'\)\), \nonumber \\
\Pi_E &=&
\frac{2a^2\sqrt{\gamma}\Delta}{\kappa}\(\psi'-\frac{\kappa}{2}\phpr\dph+\H
A-{\cal K}\(B-E'\)\), \nonumber \\
\Pi_{\dph} &=& a^2\sqrt{\gamma}\(\dph'-\phpr A\).
\ea
Under an
infinitesimal scalar coordinate transformation
$x^{\mu}\rar x^{\mu}+\lambda^{\mu}$, where
$\lambda^{\mu}=\(\lambda^0,\lambda^{|i}\)$, the perturbation fields
and momenta transform as 
\ba
&&\psi\rar\psi-{\cal H}\lambda^0,  ~~  B\rar B+\lambda'-\lambda^0, ~~
A\rar A+{\lambda^{0}}'+{\cal H}\lambda^0,  ~~  E\rar E+\lambda,
  ~~ \dph \rar \dph+\phpr \lambda^0, \nonumber\\
&&\Pi_{\psi}\rar\Pi_{\psi}+{2 a^2 \sqrt{\gamma}\over \kappa}(\delk) \lambda^0,
  ~~~~  \Pi_{E}\rar\Pi_{E},  ~~~~  \Pi_{\delta \phi} \rar\Pi_{\delta \phi}
+a^2 \sqrt{\gamma}(\phi''-{\cal H} \phi')\lambda^0.
\labeq{gts}
\ea

\section{Hawking-Moss Instantons}

Let us first
consider Hawking-Moss instantons~\cite{hm}, where the scalar
field is everywhere a constant.  It is well-known that these have
negative eigenmodes for $V_{,\phi\phi}<0$.  The corresponding
Lorentzian solution has $\phpr=0$ everywhere and so $V_{,\phi}$
must be zero.  We
notice immediately from Eq.~\eqn{2acnew} that the matter and
metric degrees freedom decouple, and from Eq.~\eqn{gts}
that $\dph$ has become a gauge invariant variable.  
Introducing
conjugate momenta to the gravitational degrees of freedom, and
performing the integrals, we  find
that no gauge invariant combination of the fields and momenta that is
not forced to be zero is left.  This means that there are no real
degrees of freedom described by the metric.  Indeed $\Psi_N$ is forced
to be zero 
here, making one wary of any approach (such
as that in~\cite{sastan,tan}) pertaining to 
negative modes that relies on metric variables alone.  Returning to the
matter degree of freedom, we analytically continue into the Euclidean region
as detailed in ~\cite{gt},
leaving us with the action
\ba
\frac{1}{2} \int dX d^3 x b^2 \sqgam \( \dph'^2 - \dph \Delta_3 \dph +
b^2 V_{,\phi\phi} \dph^2\) .
\ea
where $\gamma$ is now the determinant of the Euclidean metric,
and the Euclidean background line element is $b^2(X)(dX^2 
+\gamma_{ij}dx^i dx^j)$ with $\gamma_{ij}$ the metric
on the round three sphere. 
All gradient terms are positive so this action is bounded below for
square-integrable variations of the scalar field.  The first thing to
note is that if $V_{,\phi\phi}>0$ then this action is positive
definite and so the spacetime is perturbatively stable.  If
$V(\phi_0)$ is the global minimum of $V$ one might expect that this
spacetime is 
non-perturbatively stable as well.  We can see the existence of a
negative mode for $V_{,\phi\phi}<0$ as follows.  The eigenvalue
equation associated with this action is of Sturm-Liouville form, but
we have some freedom in specifying the measure, which we shall
repeatedly exploit.  $b^4 \sqgam$ is a
permissible choice, allowing us to just read off that $\dph=$~constant is
an eigenmode with eigenvalue $V_{,\phi\phi}$.  One might enquire if
there is another negative mode.  Rescaling $\dph$ by a factor of $b$,
the action operator takes Schr\"{o}dinger form and choosing the
measure to be $\sqgam$ gives us the Schr\"{o}dinger equation with a
$-(2-V_{,\phi\phi}/H^2 )/b^2 +n^2$ potential, where $n^2= -\Delta_3+1$, with
$\Delta_3$ the 
Laplacian on $S^3$.  This can be solved for
the negative eigenvectors and eigenvalues (see~\cite{ll} for example)
in terms of hypergeometric functions.  The number of negative modes is
independent of the choice of measure.  For $-4<V_{,\phi\phi}/H^2<0$
there is one, for $-10<V_{,\phi\phi}/H^2<-4$ there are six, and in
general for $-N(N+3)<V_{,\phi\phi}/H^2<-(N-1)(N+2)$, where $N \in
\mathcal{Z}^+$,  there are
$N(N+1)^2(N+2)/12$ negative modes.  This counting agrees with an
$O(5)$ spherical harmonic analysis.  We see that as $V_{,\phi\phi}$
becomes more negative from zero five more modes suddenly cross zero at
$V_{,\phi\phi}/H^2=-4$, meaning that the Hawking-Moss instanton cannot
now have anything to do with tunneling.  This is very interesting
because one of the new negative modes is spatially homogeneous, and
antisymmetric in $X$.  In fact it is the perturbative indication of
the existence of a lower-action non-perturbative solution, namely the
Coleman-De Luccia instanton for the same potential.
$V_{,\phi\phi}/H^2<-4$ is the precise condition for the existence of a
Coleman-De Luccia instanton~\cite{stein}, which has lower action.  We
will show below that this itself has a negative mode, which may be
viewed as the carry-over of the lowest one of the Hawking-Moss
instanton. The spectral flow is as follows. For small negative 
$V_{,\phi\phi}$, there is only one classical solution (the Hawking-Moss
instanton) with one negative mode. As $V_{,\phi\phi}$ becomes more
negative, one of the positive eigenvalues  decreases to zero. 
As it crosses zero, a new classical solution is 
obtained by flowing down the new negative direction to a 
new saddle point, which retains the original negative mode but is 
stable in other directions. 
The spatially homogeneous negative
modes gained by passing through $V_{,\phi\phi}/H^2=-N(N+3)$, $N>1$
correspond to the coming into existence of non-perturbative multibounce
instantons. From spectral flow arguments one would expect these
to possess $N$ negative modes. The same arguments indicate that
the lowest action
Coleman-De Luccia instanton should only have one negative
mode, and this is confirmed numerically as we discuss below.

\section{Coleman-De Luccia Instantons}

Let us now consider Coleman-De Luccia instantons~\cite{coldel}.  We
start in the 
open universe from Eq.~\eqn{2acnew}.  For $\Delta \neq 0$ we
proceed as in Sec.~4 of Ref.~\cite{gt} to Eq.~(20) there.  We do the
$B$  
and  
$\Pi_E$ integrals, effectively setting $\Pi_E=0$ in that
expression.  For $\Delta=0$ $B$ and $E$ 
no longer appear in Eq.~\eqn{2acnew}, and we cannot define a $\Pi_E$.
However we can still introduce $\Pi_{\psi}$ and $\Pi_{\dph}$ and work
forward to the same expression in terms of $\psi$,
$\Pi_\psi$, $\dph$, and $\Pi_{\dph}$ as for $\Delta\neq 0$.  So from
now on we treat
$\Delta=0$ and 
$\Delta\neq 0$ in a unified way. $\Psi_N$, as defined in
Eq.~(21) of
Ref.~\cite{gt} 
 is singular for $\Delta=0$, and from
our experience with the Hawking-Moss case above we know that spatially
homogeneous fluctuations are significant when investigating negative
modes.  So we define the closely related variable 
\ba  
\Psi_l= \(\delk\) \psi +\frac{\H \kappa \Pi_\psi}{2 a^2 \sqgam}
\labeq{flucpsi}
\ea
where we have also taken the opportunity to multiply through by
$\delk$ in order to keep our fields local. 
This is gauge invariant
and classically the same as $\(\delk\) \Psi_N$ since $\Pi_E$ is constrained
to be zero.  In~\cite{gt}, introducing
$\Pi_N$ made the action 
independent of $\dph$.  This is classically equivalent to working in
a gauge $\dph=0$, and from the Hawking-Moss example we see that this
is potentially 
awkward.  So here we define 
\ba
\dph_l= \(\delk\)\dph-\frac{\kappa \phpr\Pi_\psi}{2 a^2 \sqgam}
\ea
which is again local and gauge invariant.  Using $\dph_l$ and $\Psi_l$
is classically 
equivalent to working in the gauge $\Pi_\psi=0$, which from
(6) and for ${\cal K} \neq 0$ is a good gauge
everywhere.  The action now takes the simple form
\ba
\int d\tau d^3x \frac{a^2 \sqgam}{\delk} \Bigg\{ \frac{2}{\kappa \phpr}
\Psi_l \dph_l' 
+\frac{2 \(\H\phi'-\phi''\)}{\kappa \phi'^2} \Psi_l \dph_l \nonumber \\
+\frac{1}{2} \dph_l^2  
 -\frac{1}{\kappa} \(1+2 \(\delk\)/ \kappa \phi'^2\)
 \Psi_l^2\Bigg\}.  
\labeq{beaut}
\ea
It is remarkable that in the $\Pi_\psi=0$ gauge, the Newtonian potential
$\Psi$ and the scalar field fluctuation $\delta \phi$ are
the two remaining physical, and canonically conjugate variables.

We now have a choice in deciding which linear combination of
$\dph_l$ and $\Psi_l$ to retain as our coordinate before continuing
to the Euclidean region. 
Having made this choice we integrate out any remaining non-parallel
combination and obtain a quadratic action for the coordinate of
interest, which we continue to
the Euclidean 
region.  For a given background instanton, if the Euclidean action has
positive definite derivative 
terms it is bounded below for normalized square integrable fluctuations
of that 
variable, and we
have made a good choice for isolating negative modes.
From our 
experience with Hawking-Moss an obvious choice is to take $\dph_l$
itself.  After analytic continuation to the Euclidean region, 
followed by a simple rescaling $Q= b \dph_l
/\phi'$, the action takes the form  
\ba
\int \frac {dX d^3x \sqgam \phi'^2}{2 \(-\Delta_3-3\)^2}
\(\frac{Q'^2}{1+\frac{\kappa
    \phi'^2}{2\(-\Delta_3-3\)}}+\(-\Delta_3-3\)Q^2\).
\labeq{cdelact} 
\ea

Let us first briefly discuss the technicality of what happens when
$\(-\Delta_3-3\)=0$, corresponding to $n=2$ (recall that the
eigenvalues of the Laplacian on $S^3$ are $-n^2+1$, $n \in
{\cal N}$).   
This action is infinite unless $Q=0$.  One
takes this to be a positive infinity since then it 
says that $\(-\Delta_3-3\)=0$ modes are infinitely suppressed.
That this is correct can be seen by  considering Eq.~\eqn{2acnew}  for
this  mode in a  closed universe with $\phi' \neq 
0$.  There is a degeneracy between $\psi$ and $E$, resulting in only
the sum $\psi+E$ affecting the three-metric.  Then there are no
gauge-invariant combinations of fields and momenta that are not forced
to be zero.  Note that
$\(-\Delta_3-3\)=0$ modes do exist for the Hawking-Moss instanton but
in that case $\phi'=V_{,\phi}=0$ everywhere.

Having dealt with this, we move on to the more
interesting cases of $n=1$ and $n>2$.  In the latter inhomogeneous
case both the kinetic terms and the potential terms are positive
definite, giving no possibility of negative modes.  Now
let us consider the potentially 
dangerous 
homogeneous $n=1$ mode.  The kinetic term is positive definite so long
as $1-\kappa \phi'^2 /6 > 0$ across the entire instanton.  This
condition holds for a wide class of Coleman-De Luccia instantons, of 
both the thin-wall and the thick-wall variety.  In this case we see
the existence of the 
negative mode $Q=$~constant by choosing the measure $\phi'^2 \sqgam$.
So a wide 
class of Coleman-De Luccia instantons are shown to have a negative
mode.  Incidentally we note that had we
chosen a variable that had no $\dph$ matter component, the homogeneous mode
would have had negative definite kinetic term, as found
in~\cite{sastan,tan}.  Then the action could be arbitrarily
negative for square integrable fluctuations of the metric variable.

Having chosen a measure, one can numerically determine the other
eigenmodes and eigenvalues of the operator.  For Coleman-De Luccia
instantons associated with potentials of Gaussian
form $e^{- A \phi^2}$ for example, we have found no evidence of further
negative modes about these lowest-action regular solutions.
This  
is consistent with expectations based on spectral flow from the
Hawking-Moss instanton as discussed above.

\section{Hawking-Turok Instantons}

On singular instantons, as the scalar field tends to infinity, the
condition $1-\kappa \phi'^2 /6 < 0$ is certainly violated.  However
this by itself does not mean that we should exclude them. Rather we
should first consider the possibility that gravity on the 
instanton is sufficiently strong that a pure matter variable like $Q$
does not provide a suitable description of the fluctuations.  Indeed, going
back to Eq.~\eqn{beaut}, we can define $\bar{Q}=\dph_l +2
\(\H-\phi''/\phi'\) \Psi_l /\kappa \phi'$ to obtain after analytic continuation
\ba
\int \frac{ dX d^3x b^2 \sqgam}{2 \(-\Delta_3-3\)^2} 
\(\frac{\(-\Delta_3-3\) \bar{Q}'^2}{(\phpr
  (1/\phpr)''-\Delta_3-4)} +\(-\Delta_3-3\) \bar{Q}^2 \).
\labeq{htact}
\ea 
This time we see that the kinetic term is positive definite both for
$n=1$ and $n>2$ as long as $-4<\phpr (1/\phpr)''<4$.  Using the
background field equations, we have
\ba
D(X)\equiv\phpr (1/\phpr)''-4=-b^2 \(2\kappa V+\frac{8 \H
  V_{,\phi}}{\phpr}+V_{,\phi\phi}-\frac{2 b^2
  V_{,\phi}^2}{\phpr^2}\).
\ea 
We see that if the potential has a maximum then we
must have $2\kappa V+V_{,\phi\phi}>0$ for $\bar{Q}$ to be a suitable
variable.  Let us examine the behaviour of this term near the
singularity.  We have $b^2$ going like $X$, and $\phi$ goes like $-\ln
\sqrt{3/2\kappa} X$.  If $V$ is polynomial, $D$ goes like $X$
times a term involving $\ln X$ factors. Now the solution of the
eigenvalue operator for any 
eigenvalue is of
the form $A \int D(X)/X dX +B$ near the singularity and we see that this
has finite action for any $A$ and $B$.  This 
shows that the 
action alone does not in fact impose the boundary conditions for the
$O(4)$ invariant 
perturbations of singular 
instantons.  It is consistent with the
fact that singular instantons cannot be regarded as
unconstrained saddle points of the Euclidean action since the action
varies across the class of singular instantons. They must be defined
by introducing a constraint into the path integral which 
is later integrated over. This constraint determines the 
allowed $O(4)$ invariant modes. 
If one is interested in calculating a
correlator which weights particularly strongly towards a given value
for the constraint (for example if we are interested in correlating
with the observed value of $\Omega$ today), it may be useful to only
consider one sector and ignore the integration over the constraint.
This is what is effectively done in~\cite{gt,ght}
where a constraint 
is implicitly applied to give an acceptable value of $\Omega_0$, and
homogeneous fluctuations are ignored since they do not affect the
microwave background correlations.

\section{Regularized Instantons}

In the above section we saw that if $V$ were polynomial in $\phi$,
then $D$ went like $X$ times a term involving $\ln X$ factors.  
However, if $V$ is asymptotically of
the form $(e^{-\sqrt{2\kappa/3} \phi})^r$, with $r$ an odd 
integer, then $D$
goes like 
$X^{r+1}$, and the eigenfunctions have 
the form $A X^{r+1}+B$ near the singularity.  Therefore with
this form of potential 
the theory has good  analytic behaviour near the singularity.
This suggests that with this type of potential there is special 
behaviour and indeed this is 
the case. We see here the perturbative indication
of the scenario of Kirklin, Turok and Wiseman~\cite{ktw}. 
There it is shown that singular instantons of potentials with the
asymptotic form $(e^{-\sqrt{2\kappa/3} \phi})^r$, with $r$ odd
and greater than $-3$,   may be viewed
as true classical solutions of a theory
related to the original theory 
by a conformal transformation which
vanishes at the Einstein frame singularity. In terms of the
new variables the metric is strictly Riemannian and the
instantons are regular.

If we use this scheme to regularize the singular instantons occurring
in a generic inflationary theory, we must modify the potential so that
it tends to 
$(e^{-\sqrt{2\kappa/3} \phi})^r$ at large $\phi$.
The theory is then defined as the limit where this modification occurs
at infinitely large $\phi$. 
We must check that our results 
are insensitive to the details of how the limit is taken. 
We choose $r$ positive because for $r=-1$, the
function $D(X)$ vanishes making the kinetic term for the fluctuations
ill-defined.

In the regularised theory, the appropriate
degrees of freedom are combinations of the conformal factor and the
scalar field, living on a regular Riemannian 
manifold.  This manifold is taken to
have the topology of $RP^4$, and the conformal factor is taken to be
in the twisted sector.  This enforces the conformal zero,
corresponding to the singularity.  On the non-contractible
three-surface where the conformal factor is zero, one is free to 
specify the Riemannian three-metric, and this corresponds to information
stored ``at the singularity'' in the original Einstein frame.

The appropriate action in
this picture is one where the Riemannian three-metric is fixed on the
conformal zero. 
In terms of Einstein frame variables, this action is just the
standard first derivative action~\cite{dirac} including the usual 
Gibbons-Hawking boundary term. 
For $O(4)$ invariant
solutions the boundary data may be taken to be the value of $m$ on
the three-surface, $m_B$, where $m$ is the Riemannian frame
radius given by 
$m\equiv b e^{\sqrt{\kappa/6} \phi}$ in terms of Einstein 
frame variables. We treat the value of $m_B$ at the conformal
zero as a variable to be integrated over in the path 
integral. For $m_B$ smaller than some value there
is no classical solution. However, for larger $m_B$ there
are two solutions, one of higher and one of lower Euclidean action. 
The higher action solution corresponds to low values of the
scalar field $\phi_0$ at the beginning of the Lorentzian open universe. 
The lower action solution corresponds to a larger value for 
$\phi_0$. As $m_B$ is increased, the corresponding value of
$\phi_0$ increases to infinity, giving larger and larger
amounts of inflation in the Lorentzian universe.

It is slightly subtle to impose the required 
constraint because the single
field degree of freedom we use is not $\delta m$.  However we express
$\delta m$ in terms of 
$\bar{Q}$ and its canonical conjugate, as given by its saddle-point
value in the path integral.  Consider working in the gauge
$\dph=0$.  This is a good gauge near the singularity because $\phi$ is
varying quickly there.  Then $\delta m$ goes like $\psi$, which in
terms of our gauge-invariant local variables is proportional to
$\Psi_l+\H \dph_l / \phi'$.  Now $\bar{Q}=\dph_l +2
\(\H-\phi''/\phi'\) \Psi_l /\kappa \phi'$, and at the saddle point
$\Psi_l=\kappa \phi' \bar{Q}' / 2 D(X)$.  Consequently
$\delta m $ is proportional to $\bar{Q}-3 \bar{Q}' /
\H D(X)$ near the singularity.  Our numerical 
code uses the auxiliary variable $P\equiv b^2 \bar{Q}' / D(X)$ and
works in proper Euclidean time.  So the condition that we must impose on our
eigenfunctions is that $\bar{Q}+3 P / b^2 \dot{b}=0$ at the
singularity. It is straightforward to show that the most general
solution of the $\bar{Q}$ eigenvalue equation has the 
behaviour $A X^2 +B$ near $X=0$, and our boundary condition is
a specific relation between $A$ and $B$.

We have investigated a number of potentials which behave appropriately
at large $\phi$.  For example we have matched a $\phi^2$ potential onto the
$e^{-\sqrt{2\kappa/3} \phi}$ potential using a negative cubic term.
  One has to be slightly careful with the matching prescription so
  that one does not violate the $2\kappa V+V_{,\phi\phi}>0$ condition
  for $\bar{Q}$ to be a good variable at the turnover point.  As long
  as the matching is done a long way further along the potential than
 where the runaway behaviour starts, the results are in any case
  insensitive to 
  the details of the matching.

\begin{figure}
\centerline{\psfig{file=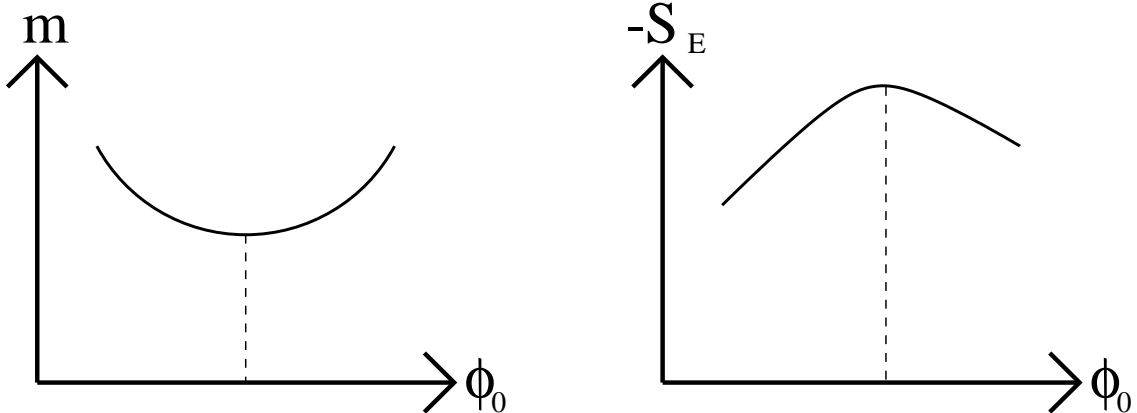,width=15cm}}
\caption{The left sketch shows how $m$ at the singularity varies with
  the value of 
  $\phi$ at the regular pole.  The right sketch shows how the
  appropriate action
  of the singular instanton varies with the value of $\phi$ at the
  regular pole.} 
\labfig{neggraph1}
\end{figure}

Now as explained above there are two starting values of $\phi$ at the regular
pole which lead to the same value of $m_B$ at the singularity (see
Fig.~\fig{neggraph1}).   
The instanton with larger $\phi$ at the regular pole has lower action.
Since for fixed $m_B$ these are the only two extrema, one could
anticipate that the larger $\phi$ solution would be stable and
the lower $\phi$ solution unstable. 
We have confirmed this numerically.

Now as we vary $m_B$ downwards the values of $\phi$ at the 
regular pole in the two solutions move closer and ultimately
merge, in the unique solution with minimal $m_B$. 
The associated instanton is the one with the most
negative action, and it is 
is like a critical point.  Since two solutions, one unstable
and one stable, are merging, one expects to find the
resulting configuration has a zero mode and this is indeed
confirmed numerically. 

As a result of this investigation we can build up a picture of the
action-configuration space  structure of the theory as shown in
Fig.~\fig{neggraph2}, and we can 
speculate as to what the structure might look like away from where we have
been able to probe.  For $m_B$ above the critical value, there
is a stable valley in $m_B$, $\phi$ space where the stability
increases with increasing $m_B$. The instantons with lower action 
are constrained solutions which lie on the floor of this valley. 
However at lower $\phi$ there is an unstable ridge, which is joined to
the valley at the critical $m_B$. The implication is that even though
the constrained instantons in the valley are stable, there are
nonperturbative instabilities lurking at low $\phi$, beyond the unstable 
ridge, and at low $m_B$, below the critical point. 
Hence it seems unlikely that the
Euclidean path integral will be well defined nonperturbatively.
It appears that
at the very least projection operators onto certain subclasses of
configuration space in the path integral are required.    
 
\begin{figure}
\centerline{\psfig{file=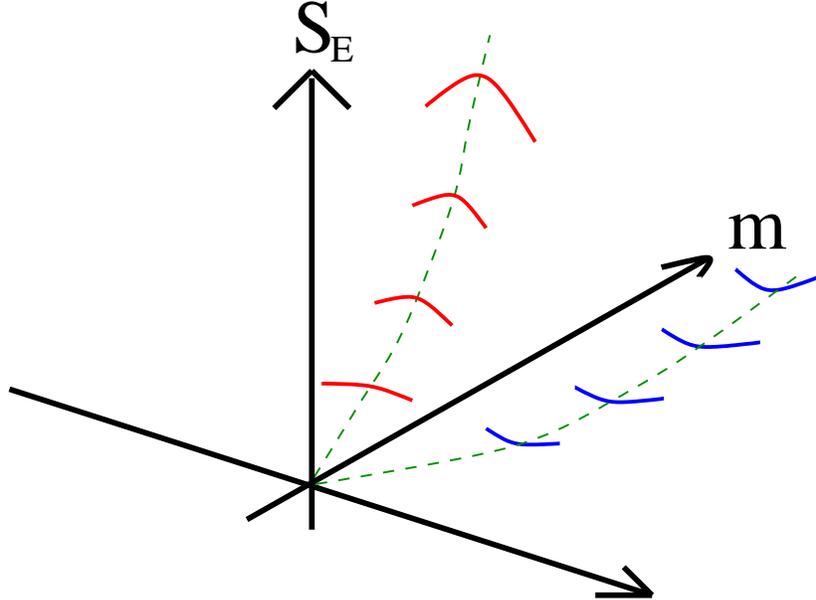,width=11cm}}
\caption{A sketch of the action-confguration space structure of
  gravity with scalar field.  The dashed lines represent the instanton
  solutions for a given $m$, and the third direction represents the
  infinity of field configurations respecting this $m$ constraint.
  The curves indicate if the instantons possess negative modes or
  not.}  
\labfig{neggraph2}
\end{figure}

\section{Connections with Previous Work}

In this section we briefly show that 
the approach presented in this paper leads to the same results as
in~\cite{gt} for the computation of CMB 
background 
anisotropies about singular instantons.  One needs to check that the
spatially inhomogeneous modes 
allowed for one choice of path integral variable correspond to the
equivalent modes allowed for the other choice of variable.  For the
inhomogeneous modes, 
$\Psi_l$ and $\Psi_N$ are equivalent, and we shall show that the $q$ modes
allowed in~\cite{gt} give the same
behaviour in $\Psi_N$ 
near the singularity as the allowed $\bar{Q}$ modes here give in $\Psi_l$.
In~\cite{gt}, the unsuppressed $q$
modes behaved as 
$X^{3/2}$, corresponding to $\Psi_N$ tending to a constant.  The
suppressed mode had $q\rar X^{-1/2}$, corresponding to  $\Psi_N$
diverging like $1/X^2$.  The
eigenvalue equation leading from~\eqn{htact} looks like $(X
\bar{Q}')'=0$ near the singularity, with general solutions of the form $A \ln
X +B$.  Substituting back into the action we find that the $\ln X$
solution 
has infinite action and so is suppressed.  At the saddle point we have
$\Psi_l$ behaving as $\bar{Q}'/X$ and we see that $\bar{Q}\rar A \ln X$
corresponds to $\Psi_l \rar 1 / X^2$, whereas the unsuppressed mode
$\bar{Q}\rar B+O(X^2)$
corresponds to $\Psi_l$ being finite.  Hence both approaches select
the same allowed modes and thus give equivalent correlators.  

No such check is necessary for the non-singular instantons because in
this case there is no boundary and all modes
are allowed.

\section{Alternate ``Five Dimensional'' Boundary Condition}

For the special potential
$V \propto e^{\sqrt{2\kappa\over 3}\phi}= n^{-1}$, where here $n
\equiv e^{-\sqrt{2\kappa\over 3}\phi}$,
Garriga showed~\cite{garriga} that 
singular Hawking-Turok instantons could be interpreted 
as ``dimensional reductions'' of a regular five dimensional 
solution, which is just a round five sphere. 
He showed that the five dimensional action, 
when written in four dimensional variables, 
differs from the 
standard first-derivative four dimensional action by minus two thirds of the 
Gibbons-Hawking
surface term. He also showed that for arbitrary potential $V$ 
one reached the same conclusion if one introduced a 
brane (of negative tension) to regularize the singularity.

In this section we study the existence of negative modes for 
a form of the action motivated by Garriga's observation, 
for arbitrary scalar potential
$V(\phi)$. Note that Garriga's five dimensional example yields
the specific potential $e^{\sqrt{2\kappa\over 3}\phi}$. 
The exact solution here has $\phi'\propto 1/{\rm sinh}2X$ and
the function $D(X)$ which enters the 
the kinetic term for the perturbations vanishes identically.
We are unable therefore
to prove existence of a negative mode in this case. Indeed this
is perfectly consistent since from 
a five dimensional
view, the Garriga solution should have no negative modes.
It is a round five sphere and continues to five dimensional
de Sitter spacetime which is presumably stable in analogy with our
treatment of the four dimensional case.  

The five dimensional line element is given in terms of the
four dimensional one $ds_4^2$ by 
$ds_5^2=n^{-1} ds_4^2
 +n^2 dy^2$ where 
$ds^2_4=
N^2 d\sigma^2+b^2(\sigma) d\Omega_3^2$  and 
$0<y\leq L$ runs around the fifth dimension,  whose radius is
${L\over 2 \pi}n ={L\over 2 \pi}  {\rm exp} (-\sqrt{2\kappa\over 3} \phi)$. 
Calculation of the
five dimensional Einstein action for gravity with a cosmological 
constant using this metric yields the action for 
four dimensional Einstein gravity plus a minimally coupled scalar 
field $\phi$ with potential $V\propto  e^{\sqrt{2\kappa\over 3} \phi} =
n^{-1}$.

The embedding in five dimensions yields a natural
regularization of the
singularity. Rewriting the line element
as $d\chi^2+m^2(\chi) d\Omega_3^2 +n^2 dy^2$, we see that the 
five dimensional metric is actually perfectly regular when 
$n$ vanishes as long as $dn/d\chi$ tends to $2\pi / L$ there, since
then the singularity is just the usual two dimensional
polar coordinate singularity
which may be removed by changing to Cartesian coordinates. 
We shall explore the consequences
of applying this boundary condition in the general case.

Setting $d \tilde{\sigma}=N d\sigma$, the four dimensional Euclidean 
Einstein-scalar field action 
is 
\be
\S_{\mathrm{Ein}}=
S_3 \int d\tilde{\sigma}
\left({1\over 2} \dot{\phi}^2 b^3 + V(\phi) b^3  -3M_{Pl}^2 b(
1-b \ddot{b}-\dot{b}^2)\right),
\ee
where $S_3= 2 \pi^2$ is the volume of the unit three sphere
and dots denote derivatives witrh respect to $\tilde{\sigma}$.
The  
last term in the integrand is $-{1\over 2\kappa} R $, with $R$ the Ricci
scalar. 
We shall be interested in rewriting this term in various ways
differing by surface terms. First we 
integrate by parts to remove the second derivatives to obtain 
the action appropriate to fixed values for
the three--metric and scalar field on the boundary, as 
discussed by Dirac~\cite{dirac},
\be
\S_{\mathrm{Dir}}= S_3 \int \left(
{1\over 2} \dot{\phi}^2 b^3 + V(\phi) b^3  -3M_{Pl}^2 b(
\dot{b}^2+1 )\right) = {\cal S}_{\mathrm{Ein}} -3M_{Pl}^2 S_3 \left[b^2 \dot{b}\right].
\ee
The last term is the Gibbons--Hawking boundary term.
Re-expressing this action in terms of the fields $m$, $n$,
and the coordinate $\chi$, 
we find
\be
{\cal S}_{\mathrm{Dir}} = 
S_3 \int d\chi m \left( V(n) n^2 m^3 -3M_{Pl}^2(
m'^2 n +mm'n'+1)\right)={\cal S}_{\mathrm{Ein}}
-{3\over 2} M_{Pl}^2 \left[m^3 n'\right].
\labeq{mact}
\ee
where here a prime denotes a derivative with respect to $\chi$ and we
have used the fact that $n=0$ on the boundary. 
This action is clearly stationary under 
variations satisfying $\delta m=0$ on the boundary.

If we instead we adopt the boundary condition suggested by Garriga's
construction, we fix $n'={2 \pi}/ L$ at the boundary. The appropriate
action is obtained from (\ref{eq:mact}) by integration by parts,
\be
\S_{n'} = S_3 \int m \left( V(n) n^2 m^3 -M_{Pl}^2
(3m'^2 n +3 -m^3 n'')\right)
= {\cal S}_{\mathrm{Ein}} -{1\over 2} M_{Pl}^2 \left[ m^3 n'\right],
\labeq{garact}
\ee
and we see that the Gibbons--Hawking term has been reduced
by a factor of three. 

For simple monotonic scalar potentials, the action appropriate
to the $\delta n'=0$ boundary condition is monotonically decreasing 
as $\phi_0$ decreases towards the potential minimum. If the potential
minimum is zero, the action for the constrained instantons 
tends to minus infinity.
This is quite different to the behaviour of the action
appropriate to the $\delta m=0$ boundary condition.
 The latter action has two solutions
at fixed $m_B$ above some minimal value. As we 
showed above the lower action solution has
no negative modes, giving us a picture of configuration  space 
in which the lower action solutions comprise a stable valley
running up towards $\phi_0 \rar \infty$. In contrast, 
since for generic potentials 
there is a unique solution for each value of 
$n'$ at the singularity~\cite{ktw}, and since we know that the
Euclidean action is
unbounded below, we might suspect that the action-configuration space
structure takes
the form of a single unstable ridge. We shall see that this
picture is indeed correct.

We need to rewrite the condition that the five dimensional metric
be regular in terms of 
our perturbation fields. To do so, we rewrite the four dimensional
line element in terms of comoving coordinate $X$ as in the previous sections.
Then we divide the last term in the five dimensional
line element by the first and take the square root. We find the
condition that as $X$ tends to zero,
$b_0^{-1}(1+A)^{-1}e^{-\sqrt{\kappa\over 6} \phi} (e^{-\sqrt{2\kappa
\over 3} \phi})'$ should tend to  $2\pi / L$
 where primes now refer to $X$ derivatives, 
$b_0(X)$ is the unperturbed scale 
factor and the scalar field now includes the infinitesimal
perturbation $\delta \phi$. It is convenient to 
pick a gauge where $\delta \phi=0$, which is possible near the
singularity. In this gauge, in order to preserve five dimensional
regularity we must have $A=0$ at the singularity.
Expressed in canonical variables, this condition becomes
$\Pi_{\delta \phi}/(a^2\phi')=0$. In the path integral, the
 $G_{00}$ constraint is imposed as a delta functional and
the condition on $\Pi_{\delta \phi}$ implies
$(\kappa {\cal H} \Pi_\psi/\phi' 
+6 {\cal K} \sqrt{\gamma} \psi/\phi')/(a^2 \phi')=0$, 
which from (\ref{eq:flucpsi})
and for the homogeneous mode
becomes after Euclideanization $\Psi_l/\phi'^2 \sim
X^2 \Psi_l =0$ at $X=0$. 

Having established the boundary condition $X^2 \Psi_l=0$ appropriate
to five dimensional regularity, we translate this into a boundary
condition for the 
fluctuation variable $\bar{Q}$ appropriate to the negative
mode computation. We find that  $X^2 \Psi_l \sim \kappa X^2 \phi'
\bar{Q}'/2 D(X) \sim X^{-1} \bar{Q}'$. For the regularized
instantons, the 
general solution for the mode equation for $\bar{Q}$ is specified
by its expansion $\bar{Q} \sim AX^2+B $ near $X=0$, and the 
boundary condition therefore reads 
$A=0$. 
It is easy to see that a negative mode always exists for this boundary 
condition. From Eq.~(\ref{eq:htact}) and taking the measure to be $b^2
\sqgam$, if we set $\bar{Q}= $ constant,
the action is negative. The ansatz clearly satisfies the boundary 
condition. Therefore there is at least one negative mode. 
From a numerical study, we find that for a simple quadratic 
potential, regularized at large $\phi$ as above, there is in fact only
one negative mode. 

To summarise, the condition of five dimensional regularity
may be imposed as a boundary condition. However, it
does not eliminate the negative modes 
therefore leaving the Euclidean path integral as 
ill-defined at a fundamental level.

\section{Conclusions}

In this paper we have given a detailed investigation of
spatially homogeneous fluctuations of cosmological instantons.  We
showed how a Hamiltonian treatment, with an
appropriate choice of variable, produces a 
Euclidean action which is bounded below for normalized 
fluctuations. First, we investigated 
Hawking-Moss and Coleman-De Luccia
instantons, and showed that the lowest action solution in each
case possess a negative mode. 
For the ``thin-wall'' Coleman-De Luccia case
where the instantons ``almost'' interpolate between the true and false
vacua, this 
supports their use in tunneling roles as discussed
in~\cite{calcol,rubsib}.  In that approach it is necessary that the
potential may be obtained by analytically distorting one for which the
action is positive definite for all configurations.  In the ``thick-wall''
case, and certainly for Hawking-Moss instantons, where gravitational
effects are important, it is not clear that this is possible.
Thus even though we have found that these instantons do possess a
single 
negative mode, the assertion that these instantons are useful
for describing the decay of one 
spacetime to another requires further understanding.

Our investigation of singular instantons indicates the 
importance of a well defined regularization. 
In contrast to the situation for the inhomogeneous modes~\cite{gt},
the Euclidean action does
not uniquely select a boundary condition for the homogeneous modes. 
It is therefore essential to choose a regularization
within which the relevant boundary condition is defined. 
We have investigated two such frameworks. The first is the
$RP^4$
construction of~\cite{ktw}, according to which we find
that instantons with large starting values $\phi_0$
of the inflaton field have no negative modes to quadratic order. 
The second is the regularization motivated by Garriga's 
five dimensional construction. Here we find that 
a negative mode is always present.

Since in all the cases studied here the instantons have perturbations
which decrease their action,
their use in an unconstrained path integral
to determine the quantum state of the
universe is questionable.  In the case of the instantons describing
tunnelling, a constraint is needed to set the system in an
initial unstable state. The constraints introduced in
the $RP^4$ construction remove negative modes perturbatively,
but as we have argued, probably do not remove them 
nonperturbatively. It therefore seems essential that an
additional constraint be introduced which effectively projects
onto certain subsets of states, and excludes the 
configurations of arbitrarily negative Euclidean action.
This might be justified if we are only interested in
correlators of certain subsets of observables for example,
as opposed to the unconstrained Euclidean partition function.
How the appropriate 
projections are to be defined and introduced is an
important topic for future work.

\medskip
\centerline{\bf Acknowledgements}

We wish to thank S. Hawking, 
K. Kirklin and T. Wiseman for useful discussions.

This work was supported by a PPARC (UK) rolling grant and a PPARC studentship.

\end{document}